# Equilibria, periodic orbits around equilibria, and heteroclinic connections in the gravity field of a rotating homogeneous cube

Xiaodong Liu[1], Hexi Baoyin[2], and Xingrui Ma[3]

*School of Aerospace, Tsinghua University, Beijing 100084 CHINA*

Email: *liu-xd08@mails.tsinghua.edu.cn*; *baoyin@tsinghua.edu.cn*; *maxr@spacechina.com*

## Abstract

This paper investigates the dynamics of a particle orbiting around a rotating homogeneous cube, and shows fruitful results that have implications for examining the dynamics of orbits around non-spherical celestial bodies. This study can be considered as an extension of previous research work on the dynamics of orbits around simple shaped bodies, including a straight segment, a circular ring, an annulus disk, and simple planar plates with backgrounds in celestial mechanics. In the synodic reference frame, the model of a rotating cube is established, the equilibria are calculated, and their linear stabilities are determined. Periodic orbits around the equilibria are computed using the traditional differential correction method, and their stabilities are determined by the eigenvalues of the monodromy matrix. The existence of homoclinic and heteroclinic orbits connecting periodic orbits around the equilibria is examined and proved numerically in order to understand the global orbit

1 PhD candidate, School of Aerospace, Tsinghua University

2 Associate Professor, School of Aerospace, Tsinghua University

3 Professor, School of Aerospace, Tsinghua University

structure of the system. This study contributes to the investigation of irregular shaped celestial bodies that can be divided into a set of cubes.

**Keywords** *Two-body problems • Cube • Equilibria • Periodic orbits • Homoclinic orbit • Heteroclinic orbit • Stability • Gravity*

## 1. Introduction

It was mentioned in the previous literature that to study orbits around the Platonic solids might be fruitful (Werner 1994). The cube, which is a simple one of the five Platonic solids, is the subject of this paper. The dynamics of a particle orbiting around a rotating homogeneous cube are complex despite the cube's simple shape. The existence of ring-type bounded motion in an isolated system consisting of a massive point particle and a homogeneous cube was demonstrated previously (Michalodimitrakis and Bozis 1985).

This study can be considered as an extension of previous research concerning the cases of simple shaped massive bodies. For fixed massive bodies, dynamics around a straight segment (Riaguas et al. 1999; Arribas and Elipe 2001), a solid ring (Broucke and Elipe 2005; Azevêdo et al. 2007; Azevêdo and Ontaneda 2007; Fukushima 2010), a homogeneous annulus disk (Alberti and Vidal 2007; Fukushima 2010), and simple planar plates including square and triangular plates (Blesa 2006) were investigated. While for dynamics around a rotating segment, nonlinear stability of the equilibria was determined (Riaguas et al. 2001; Elipe and Riaguas 2003), and the invariant manifolds along with periodic solutions were analyzed (Gutiérrez–Romero et al. 2004; Palacián et al. 2006). However, in these studies, the massive bodies are limited to one- or two-dimensional cases. In contrast, the current study extends the simple shaped bodies to three dimensions.

In celestial mechanics, the traditional method for representing the potential of a celestial body is to expand it into a series of spherical harmonics, which is effective for spheroid-like bodies. However, for irregular shaped celestial bodies, such as asteroids and comet nuclei, this method usually fails to converge. To overcome the drawbacks of the harmonic expansion, the mascon approximation (Geissler et al. 1996; Scheeres et al. 1998) and the polyhedral approach (Werner 1994; Werner and Scheeres 1997) were developed. It is also advantageous to divide irregular shaped celestial bodies into a finite number of simple shaped basic units, such as cubes or tetrahedra. In this way, the potential of the celestial bodies can be obtained by adding the potentials of all the cubes, especially for irregular shaped bodies, so this paper contributes to the investigation of dynamics of orbits around them. The gravity field of 433 Eros was once modeled as the summation of the gravity field of a set of tetrahedra (Scheeres et al. 2000).

In addition, this paper has implications for the study of dynamics of orbits around non-spherical celestial bodies. The investigations of the equilibria and periodic orbits around them in the gravity field of a rotating cube in the synodic reference frame provide valuable information regarding stationary orbits around non-spherical celestial bodies. Homoclinic orbits and heteroclinic orbits around the cube also offer references for orbit transfer around a certain non-spherical celestial body.

In this study, the dynamics of orbits around a rotating homogeneous cube are investigated. In the synodic reference frame, the equilibria and their linear stabilities are determined. Periodic orbits around the equilibria are also computed using the traditional differential correction method, and their stabilities are examined. Finally, in order to understand the global orbit structure of the system, the existence of homoclinic and heteroclinic orbits connecting periodic orbits is proved numerically.

## 2. The gravitational potential of a fixed homogeneous cube

Considering a homogeneous cube with edge length 2*a* and constant mass density *σ*, the potential at a certain point *P* is given by the volume integral

$$U = G\sigma \iiint_V \frac{1}{r} \mathrm{d}V \tag{1}$$

where *G* is the gravitational constant, and *r* is the distance from the point *P* to the differential mass of the cube.

Based on the polyhedral method (Werner and Scheeres 1997), the potential of the cube at the point *P* is obtained. The result is provided in the Appendix. It can be seen that the potential at the point *P* depends on cubic density, cubic edge length, and the coordinate of *P*. Compared to the finite straight segment's potential that only contains the logarithmic function (Riaguas et al. 1999), the cubic potential is much more complex, and it includes coupling terms of logarithmic and polynomial functions, as well as coupling terms of arctan and polynomial functions. It is evident that the potential is symmetrical with respect to the planes passing through the center, and parallel to the face as well as to the diagonal planes of the cube.

### 3. Equilibria of a rotating cube

It is assumed that the cube rotates uniformly around one symmetry axis with constant angular velocity $\omega$. The rotating coordinate system *Oxyz* is established with the origin *O* located at the center of the cube, the three coordinate axes coinciding with the symmetry axes of the cube, and the *z*-axis taken as the rotation axis.

In the rotating frame of reference, given the particle *P* that is initially in an orbit tangent to the *xy*-plane, then the motion of *P* will stay in the *xy*-plane due to the symmetrical characteristic of the potential. It is evident that the conservative dynamics system only has two degrees of freedom, and the motion of the particle in the *z* direction is always zero. In mechanics, an effective potential is defined as

$$W = U + \frac{1}{2}\omega^2\left(x^2 + y^2\right) \tag{2}$$

Then, the Lagrangian of the motion is stated as

$$L = \frac{1}{2}\left(\dot{x}^2 + \dot{y}^2\right) + \omega\left(x\dot{y} - \dot{x}y\right) + W \tag{3}$$

Scaling is performed such that the half-length of the edge *a* is the unit of length and $1/\omega$ is the unit of time. After scaling, the Lagrangian is rewritten as

$$L = \omega^2 a^2\left[\frac{1}{2}\left(\dot{x}^2 + \dot{y}^2\right) + \left(x\dot{y} - \dot{x}y\right) + \frac{1}{2}\left(x^2 + y^2\right) + RV\right] \tag{4}$$

where $R = \frac{G\sigma}{\omega^2 a^2}$ and $V = \frac{U}{G\sigma}$. The dimensionless parameter *R* is the ratio of the gravitational acceleration to centrifugal acceleration. The case *R*=1 is used in this study. In the rotating frame of reference, the motion equation of a particle in space in the gravity field of the rotating cube is given as

$$\begin{aligned} \ddot{x} - 2\dot{y} &= x + RV_x \\ \ddot{y} + 2\dot{x} &= y + RV_y \end{aligned} \tag{5}$$

It can be seen that the Hamiltonian System (5) is independent of time, so it has an energy integral, which can be written in the form

$$C = \frac{1}{2}\left(\dot{x}^2 + \dot{y}^2\right) - W \tag{6}$$

The equality $C = -W$ is known as zero velocity curves, which is the bound motion of the particle. Note that the variables $x$ and $y$ and the time in Eqs. (4), (5) and (6) are not the same as in Eqs. (1) and (2).

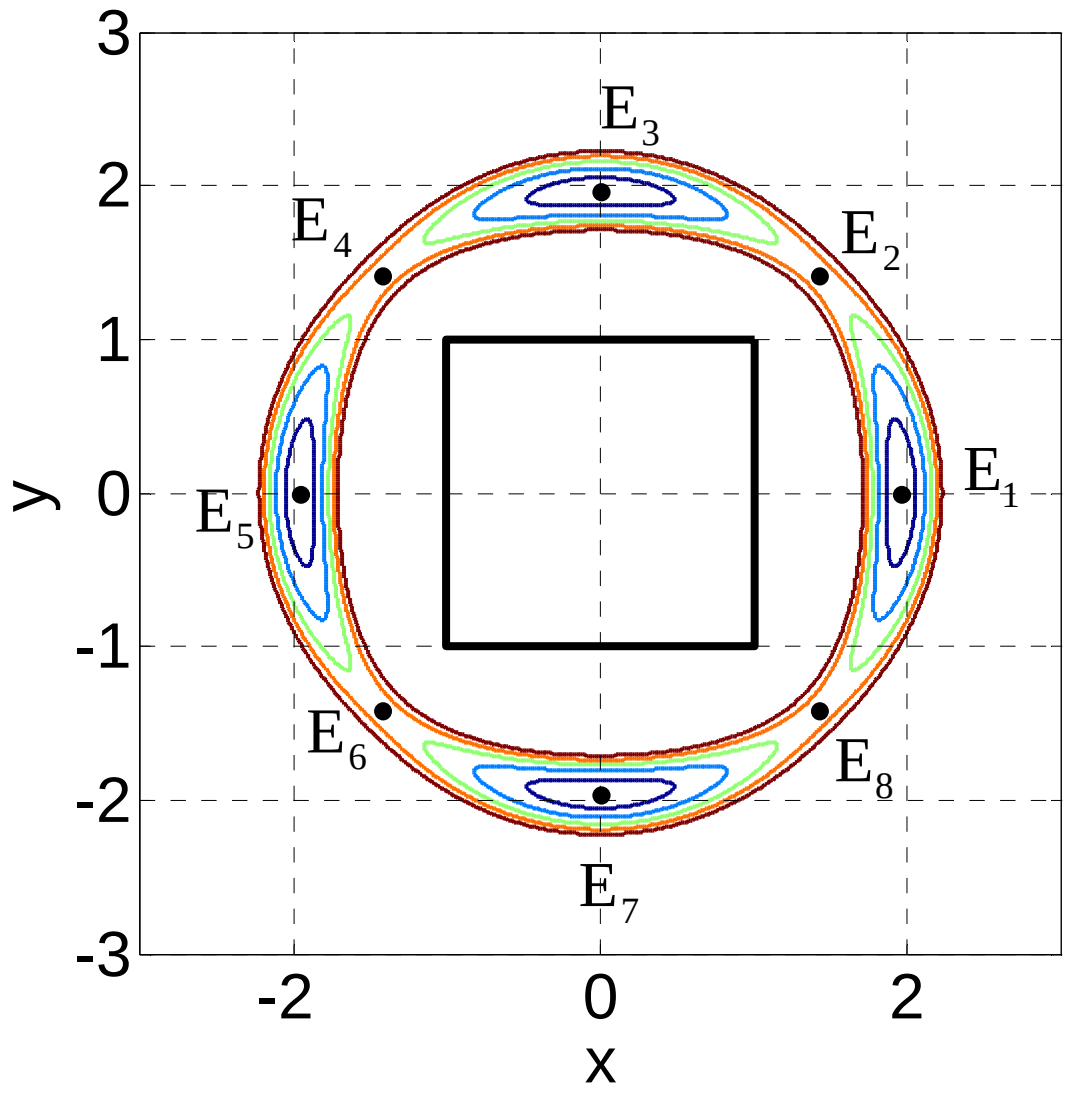

**Fig. 1** The contours of the effective potential $W$ and the equilibria in the $xy$-plane.

As shown in Fig. 1, there are eight extreme points of the effective potential, corresponding to eight equilibria of System (5) in the rotating frame of reference. It is evident that the equilibria are symmetrical with respect to the $x$-axis, the $y$-axis, and the lines $y$=$x$ and $y$=-$x$. The equilibria can be located by setting the right-hand sides of System (5) to zero, i.e.

$$\begin{aligned} x + RV_x &= 0 \\ y + RV_y &= 0 \end{aligned} \tag{7}$$

The approximate solutions displayed in Fig. 1 can serve as initial values for numerical methods. After numerical corrections, the exact values of the roots of Eq. (7) can be obtained, which are shown in Table 1. The equilibria in the rotating frame of reference can be considered as stationary orbits in the inertial frame of reference.

Based on the projection of the effective potential onto the position place, the regions of possible motion can be obtained. Let $C_i$ be the energy integral of a particle at rest at $E_i$. The evolution of the regions of possible motion for three values of the energy integral is shown in Fig. 2.

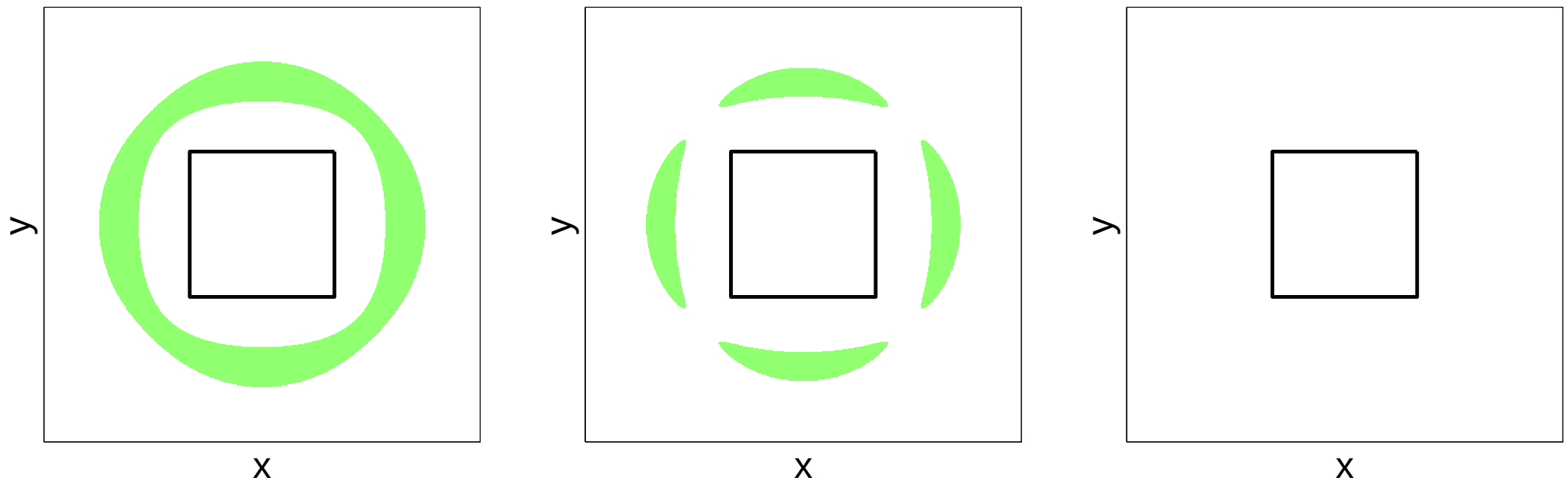

**(a)** **(b)** **(c)**

**Fig. 2** Evolution of the regions of possible motion for three values of the energy integral. (**a**) $C<C_2$. (**b**) $C_2<C<C_1$. (**c**) $C>C_1$. The shadow zones represent the regions that are inaccessible. Zones inside the shadow are known as the *interior realm*, and zones outside the shadow are known as the *exterior realm*.

**Table 1** Location of the equilibria in the rotating frame

| Equilibrium | $x$ | $y$ | Linear stability |
|---|---|---|---|
| $E_1$ | 1.958356489337404 | 0 | LS |
| $E_2$ | 1.417897298074648 | 1.417897298074648 | U |
| $E_3$ | 0 | 1.958356489337404 | LS |
| $E_4$ | -1.417897298074648 | 1.417897298074648 | U |
| $E_5$ | -1.958356489337404 | 0 | LS |
| $E_6$ | -1.417897298074648 | -1.417897298074648 | U |
| $E_7$ | 0 | -1.958356489337404 | LS |
| $E_8$ | 1.417897298074648 | -1.417897298074648 | U |

## 4. The linear stability of the equilibria

The linear stability of the equilibria can be determined by analyzing the linearized equations of System (5) near the equilibria. Because of the symmetry of the system, the stabilities of the equilibria $E_1$, $E_3$, $E_5$, and $E_7$ are the same, and the stabilities of the equilibria $E_2$, $E_4$, $E_6$, and $E_8$ are the same.

Using the notation

$$\xi = x - x_e,\ \ \eta = y - y_e \tag{8}$$

and

$$\mathbf{X} = (\xi, \eta, \dot{\xi}, \dot{\eta})^T \tag{9}$$

where $x_e$ and $y_e$ are the coordinates of the equilibrium, the linearized equations of System (5) can be expressed as

$$\dot{\mathbf{X}} = \mathbf{A} \bullet \mathbf{X} \tag{10}$$

where

$$\mathbf{A} = \begin{bmatrix} 0 & 0 & 1 & 0 \\ 0 & 0 & 0 & 1 \\ W_{xx} & W_{xy} & 0 & 2 \\ W_{yx} & W_{yy} & -2 & 0 \end{bmatrix} \tag{11}$$

The value of $\nabla\nabla V$ must be calculated at the corresponding equilibrium.

The characteristic equation of the matrix **A** can be easily obtained:

$$\lambda^4 + \left(4 - W_{xx} - W_{yy}\right)\lambda^2 + \left(W_{xx}W_{yy} - W_{xy}W_{yx}\right) = 0 \tag{12}$$

and the discriminant of Eq. (12) is

$$\Delta = \left(4 - W_{xx} - W_{yy}\right)^2 - 4\left(W_{xx}W_{yy} - W_{xy}W_{yx}\right) \tag{13}$$

For the equilibria $E_1$, $E_3$, $E_5$, and $E_7$, the discriminant $\Delta > 0$ and the four eigenvalues of Matrix (11) are calculated as

$$\lambda_{12} = \pm 0.697461937615261\, i,\ \ \lambda_{34} = \pm 0.788894954683585\, i \tag{14}$$

They are all pure imaginary, so the four equilibria are linearly stable. For the equilibria $E_2$, $E_4$, $E_6$, and $E_8$, the discriminant $\Delta > 0$ and the four eigenvalues of Matrix (11) are calculated as

$$\lambda'_{12} = \pm 1.186920914217552,\ \ \lambda'_{34} = \pm 0.544945222043182\, i \tag{15}$$

Since $\lambda'_1 > 0$, the four equilibria are linearly unstable, so they are also unstable. The states of stability of all these equilibria appear in the fourth column of Table 1, and each orbit is denoted by U (unstable) or LS (linearly stable).

Only the linear stability of the equilibria is included in this paper. The study of the nonlinear stability of the equilibria is cumbersome due to the complexity of the cubic potential, and this work is in progress.

## 5. Periodic orbits around the equilibria

Around the equilibria $E_1$, $E_3$, $E_5$, and $E_7$, the general solutions of linearized System (10) are

$$\begin{aligned} \xi &= A_1 \sin(\lambda_1 t + \varphi_1) + A_2 \sin(\lambda_3 t + \varphi_2) \\ \eta &= \alpha A_1 \cos(\lambda_1 t + \varphi_1) + \alpha A_2 \cos(\lambda_3 t + \varphi_2) \end{aligned} \tag{16}$$

where $\alpha = \frac{1}{2}(\lambda_3 + W_{xx}/\lambda_3)$ . The solutions of the linearized system can be considered as the approximate initial conditions of the periodic orbits around the equilibria. Here, only the case of $E_1$ is considered because the cases of $E_1$, $E_3$, $E_5$, and $E_7$ are symmetrical.

If $A_2$=0, $\xi = A_1 \sin(\lambda_1 t + \varphi_1), \eta = \alpha A_1 \cos(\lambda_1 t + \varphi_1)$ . Because System (5) is symmetrical with respect to the $x$-axis, the approximate initial condition of the periodic orbit is chosen as the point that intersects the $x$-axis perpendicularly:

$$x_0 = x_e + A_1,\ y_0 = 0,\ \dot{x}_0 = 0,\ \dot{y}_0 = -\alpha\lambda_1 A_1,\ T_0 = 2\pi/\lambda_1 \tag{17}$$

The initial values can be adjusted slightly by the traditional differential correction process. For example, given the amplitude $A_1$=0.1, the periodic condition is derived as

$$x_0 = 2.058356489337404,\ y_0 = 0,\ \dot{x}_0 = 0,\ \dot{y}_0 = -0.159513019894778$$

$$T = 9.331804812511473$$

Following a similar procedure, if $A_1$=0 and $A_2$=0.1, the exact initial condition of the periodic orbit is derived as

$$x_0 = 2.058356489337404,\ y_0 = 0,\ \dot{x}_0 = 0,\ \dot{y}_0 = -0.169448006096586$$

$$T = 7.802537780450530$$

Figure 3 shows the two different types of orbits around the equilibrium $E_1$, and it can be seen that both orbits are quasi-elliptical.

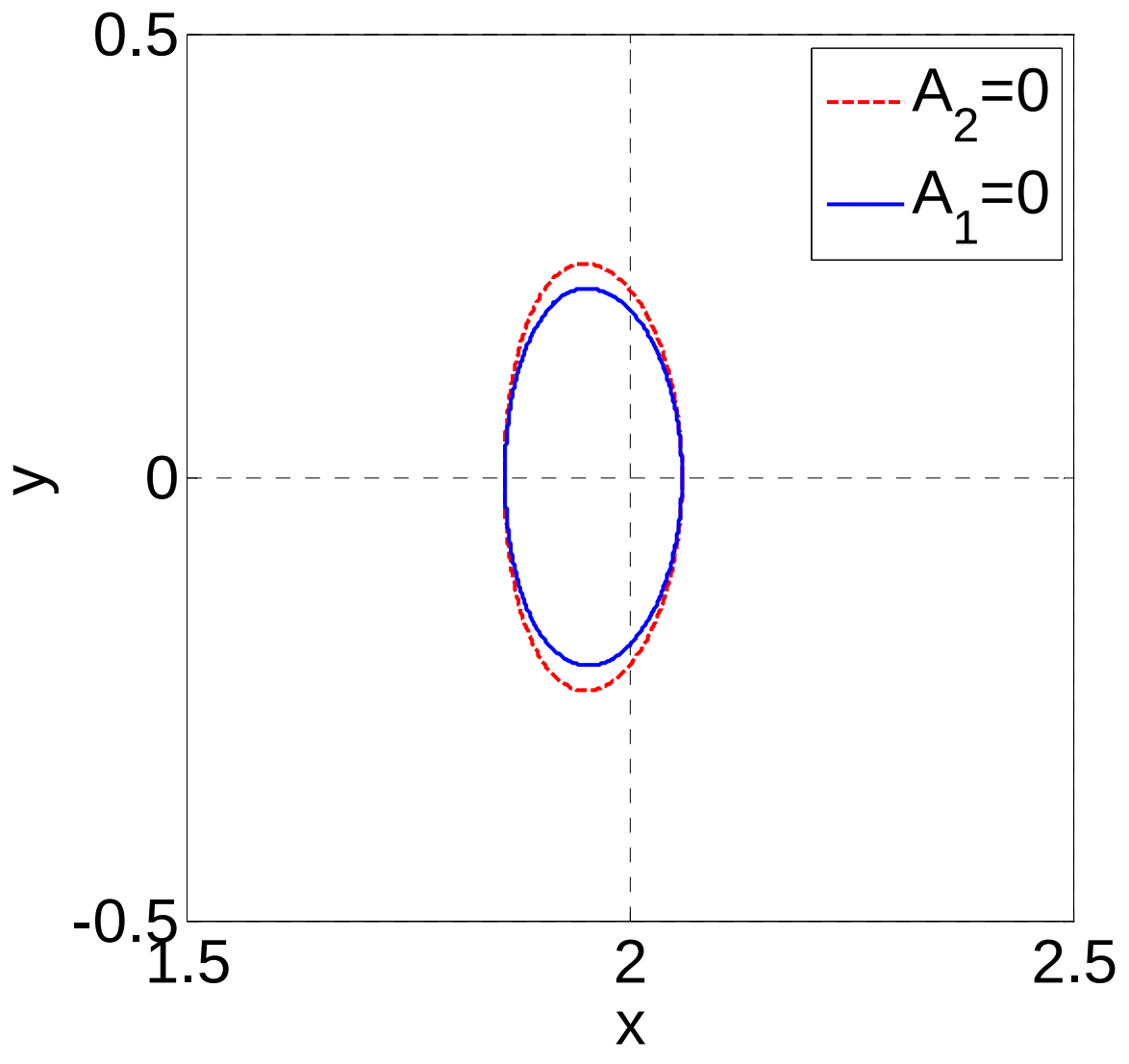

**Fig. 3** Periodic orbits around the equilibrium $E_1$. Dashed line in red corresponds to $A_1$=0.1 and $A_2$=0, and solid line in blue corresponds to $A_1$=0 and $A_2$=0.1.

The stability of the periodic orbit is determined by the eigenvalues of the monodromy matrix, and the stability index $k$ is defined as the Trace-2 of the monodromy matrix. If $k$>2 or $k$<-2, the orbit is unstable; if -2<$k$<2, the orbit is stable; and the conditions $k=\pm 2$ correspond to critical cases. Figure 4(a) shows the evolution of the stability index $k$ as a function of $A_1$ at the range of (0, 0.2] when $A_2$=0. Figure 4(b) shows the evolution of the stability index $k$ as a function of $A_2$ at the range of (0, 0.2] when $A_1$=0. It can be seen that periodic orbits around $E_1$ are stable when $A_1(A_2) \in (0, 0.2]$.

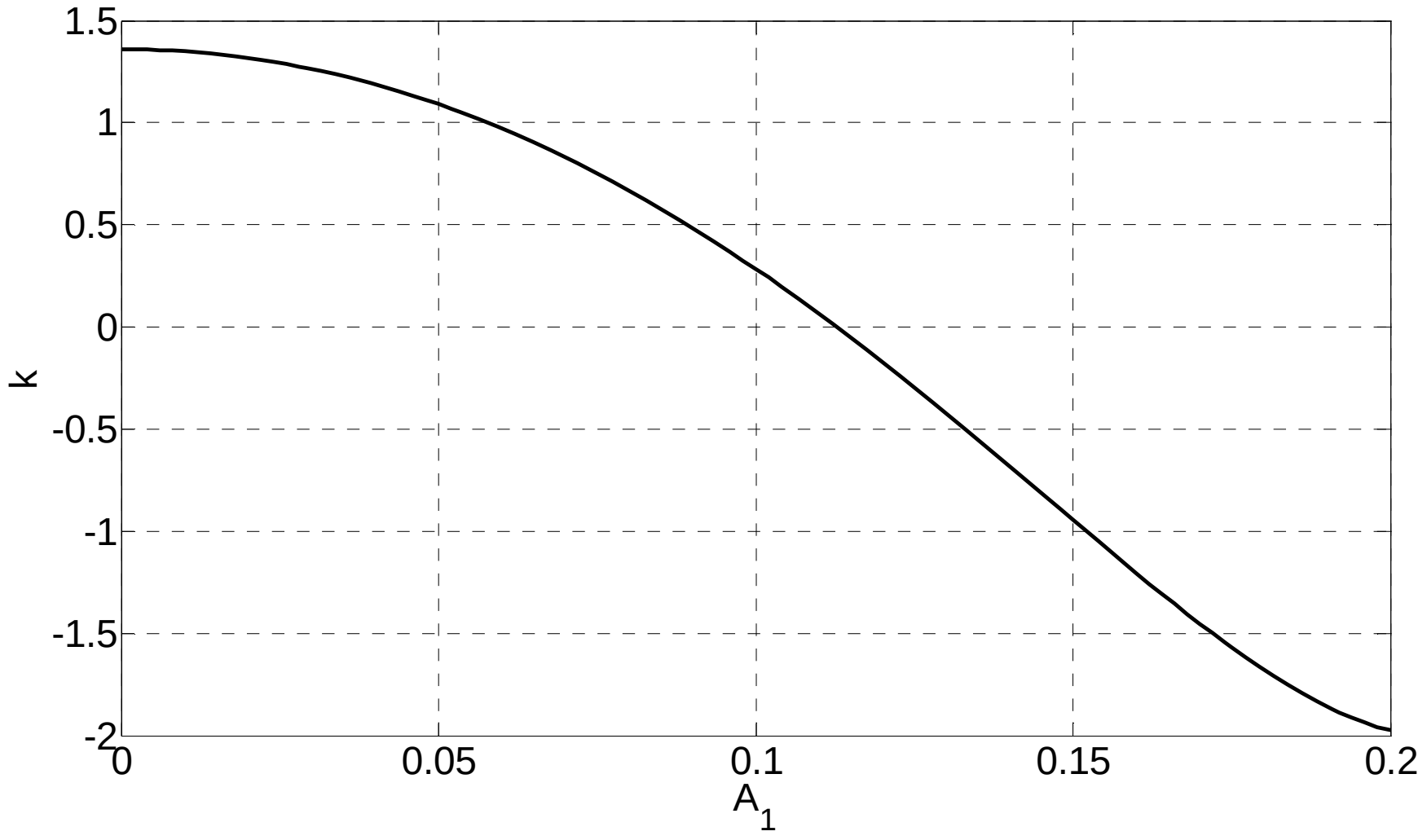

(**a**)

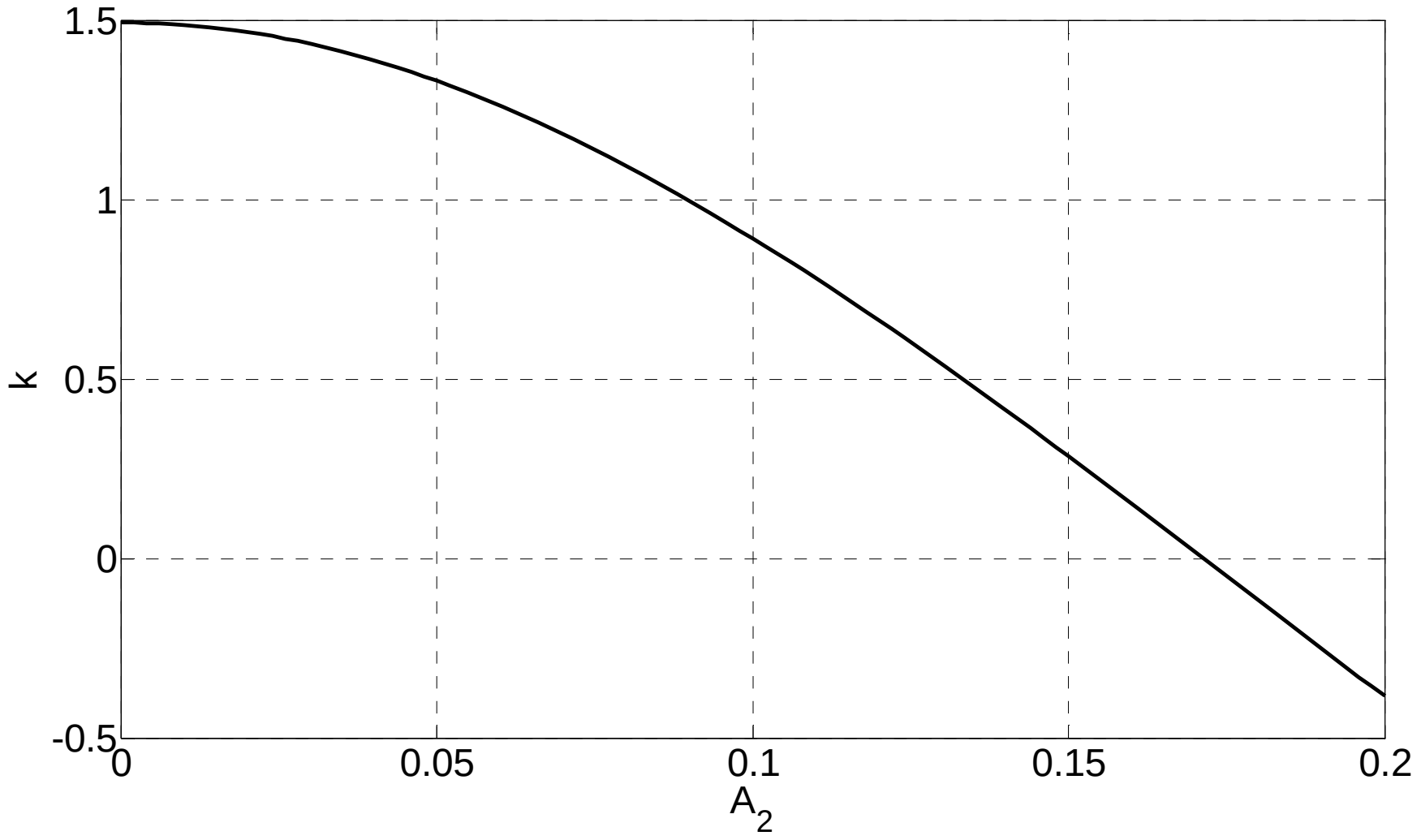

(**b**)

**Fig. 4** (**a**) Evolution of the stability index $k$ as a function of $A_1$ when $A_2$=0. (**b**) Evolution of the stability index $k$ as a function of $A_2$ when $A_1$=0.

Around the equilibria $E_2$, $E_4$, $E_6$, and $E_8$, the cases are a little different. The general solutions of linearized System (10) are

$$\xi = D_1 e^{\lambda_1' t} + D_2 e^{-\lambda_1' t} + A' \sin\left(\lambda_3' t + \varphi'\right)$$
$$\eta = \alpha_1'\left(D_1 e^{\lambda_1' t} - D_2 e^{-\lambda_1' t}\right) + \alpha_2' A' \cos\left(\lambda_3' t + \varphi'\right) \tag{18}$$

where $\alpha_1' = \frac{1}{2}\left(\lambda_1' - W_{xx}/\lambda_1'\right)$ and $\alpha_2' = \frac{1}{2}\left(\lambda_3' + W_{xx}/\lambda_3'\right)$.

In order to derive periodic orbits, the values of $D_1$ and $D_2$ must be set equal to zero, so $\xi = A'\sin\left(\lambda_3' t + \varphi'\right), \eta = \alpha_2' A' \cos\left(\lambda_3' t + \varphi'\right)$. Based on the differential correction process, the approximate initial conditions of the periodic orbits can be adjusted slightly. Taking the case of the amplitude $A' = 0.1$ as an example, the exact initial condition of the periodic orbit is obtained as

$$x_0 = 1.488607976193302,\ y_0 = 1.488607976193302$$

$$\dot{x}_0 = 0.155838488002003,\ \dot{y}_0 = -0.155838488002003,\ T = 5.293684888282230$$

Figure 5 shows that the periodic orbit around the equilibrium $E_2$ is quasi-elliptical. Figure 6 shows the evolution of the stability index *k* as a function of $A'$ at the range of (0, 0.2]. It can be seen that periodic orbits around the equilibria $E_2$, $E_4$, $E_6$, and $E_8$ are all unstable when $A' \in \left(0, 0.2\right]$.

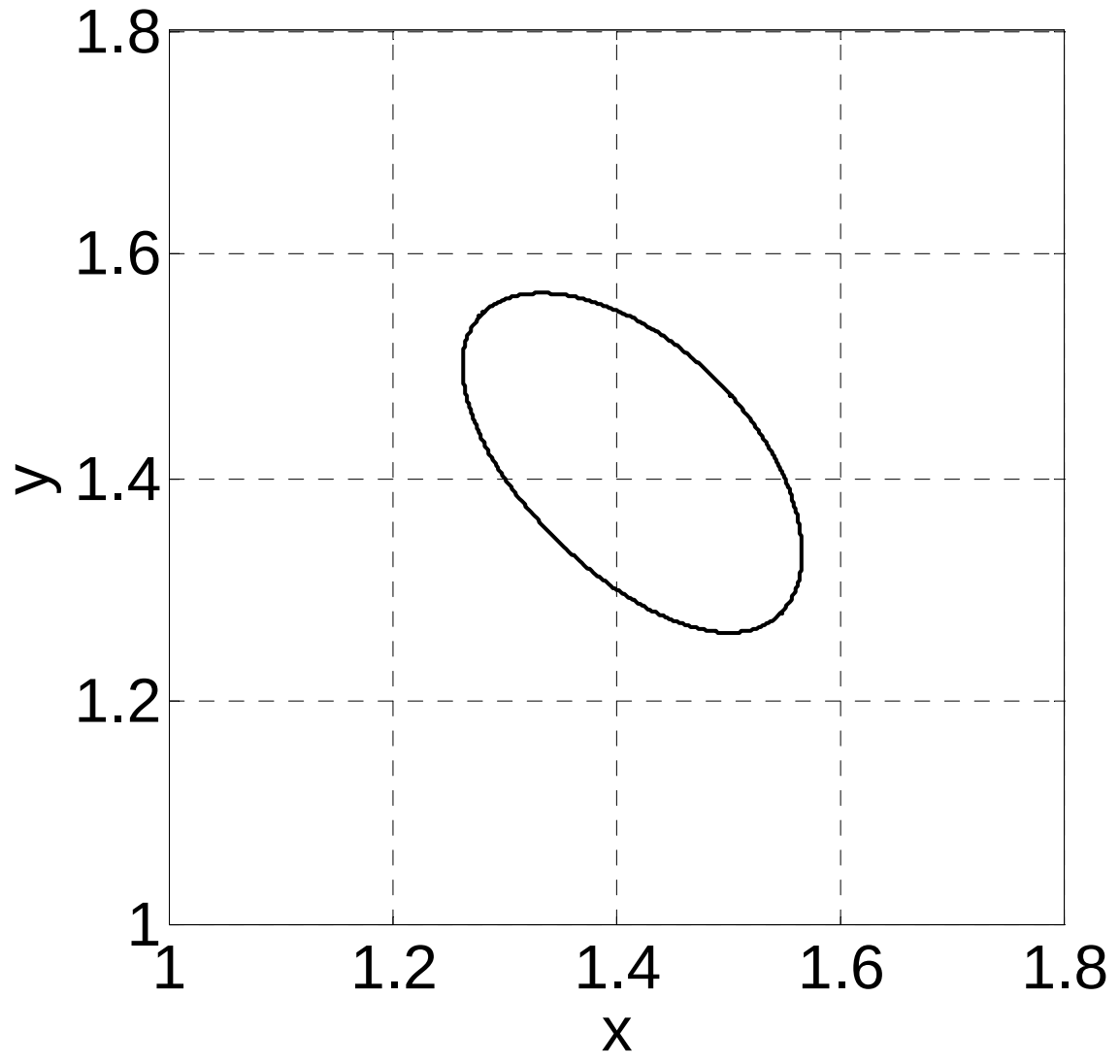

**Fig. 5** Periodic orbit around the equilibrium $E_2$ with the amplitude $A' = 0.1$.

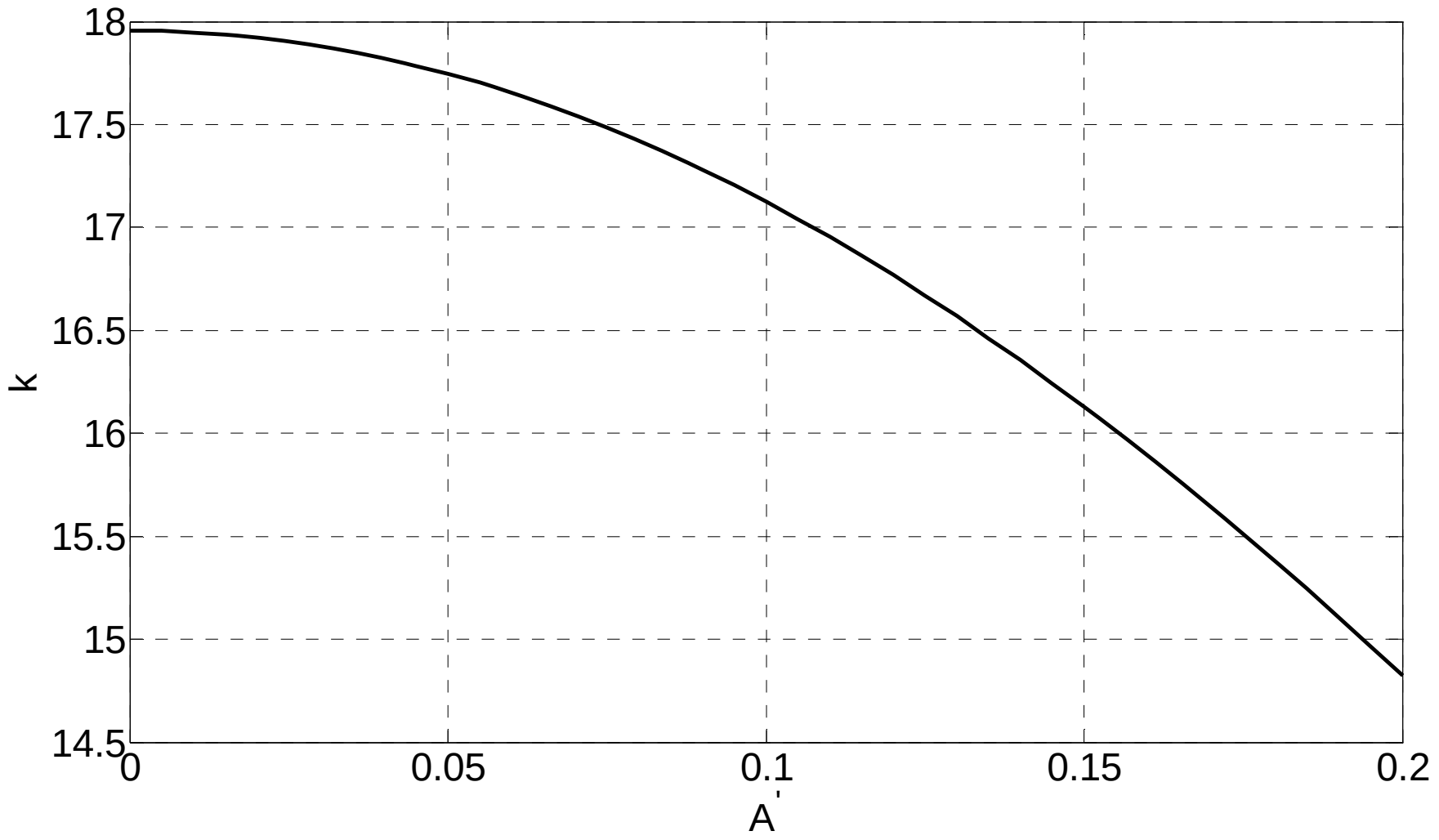

**Fig. 6** Evolution of the stability index $k$ as a function of $A'$ around the equilibria $E_2$, $E_4$, $E_6$, and $E_8$.

## 6. Homoclinic and heteroclinic orbits connecting periodic orbits

### 6.1 Homoclinic orbits connecting periodic orbits in the exterior realm ($E_2$—$E_2$)

In order to understand the global orbit structure of the system, homoclinic and heteroclinic orbits connecting periodic orbits are examined. Numerical explorations in this section prove that homoclinic and heteroclinic orbits connecting periodic orbits around the equilibria $E_2$, $E_4$, $E_6$, and $E_8$ are indeed present for the rotating cube.

At the beginning, homoclinic orbits connecting periodic orbits in the exterior realm are discussed. A homoclinic orbit connecting a periodic orbit *po* is defined as an orbit that tends toward *po* when $t \to \pm\infty$. In this case, the stable and unstable manifolds of the periodic orbit *po* coincide with each other.

Given the amplitude $A_m$=0.07, there exists a periodic orbit around $E_2$. Associated to the $E_2$ periodic orbit, there is a two-dimensional variant stable manifold $M^s_{E_2}$ and an unstable manifold $M^u_{E_2}$. Another Cartesian coordinate system $Ox_1y_1z_1$ is established with three coordinate axes $\mathbf{i}_1$, $\mathbf{j}_1$, and $\mathbf{k}_1$ parallel to **i+j**, **i-j**, and **k**, respectively. Taking the plane $y_1$ =0, $x_1$<0 as the Poincaré surface of section, the homoclinic orbit exists if the manifolds $M^s_{E_2}$ and $M^u_{E_2}$ have an intersection. In this paper, only the first intersection of $M^s_{E_2}$ and $M^u_{E_2}$ is considered because the multi-intersections are quite complicated and can appear confusing. The first Poincaré cuts of the manifolds $M^s_{E_2}$ and $M^u_{E_2}$ are denoted as $C^s_{E_2}$ and $C^u_{E_2}$, respectively.

It is evident that the new variables $x_1$, $y_1$, and $z_1$ still satisfy Eq. (5), so the system is invariant to the transformations

$$\left(x_1, y_1, \dot{x}_1, \dot{y}_1, t\right) \rightarrow \left(x_1, -y_1, -\dot{x}_1, \dot{y}_1, -t\right) \tag{19}$$

Therefore, if the stable cut $C^s_{E_2}$ is plotted, the unstable cut $C^u_{E_2}$ is obtained as the mirror image of $C^s_{E_2}$ because of their symmetry.

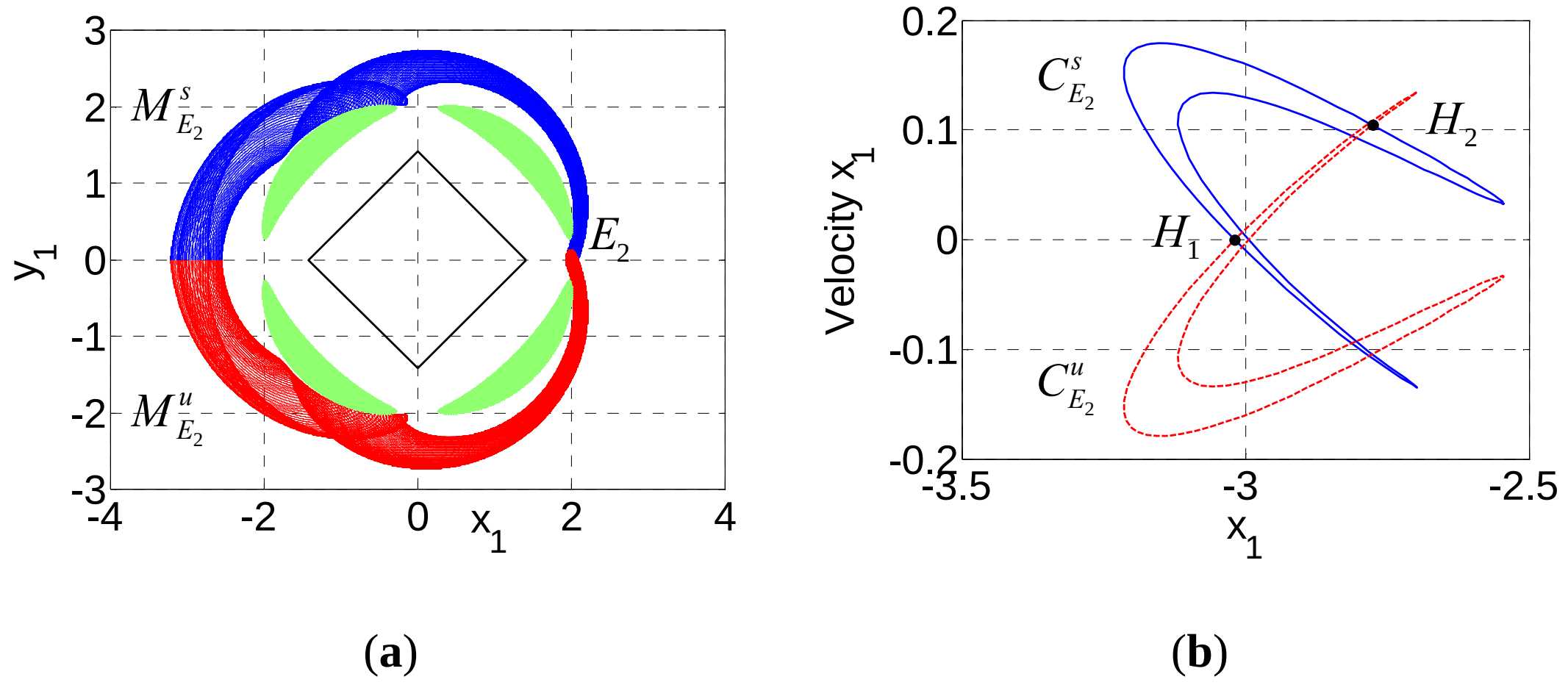

**Fig. 7** (**a**) Projection of the invariant manifold tubes $M^s_{E_2}$ and $M^u_{E_2}$ onto the position space ($x_1$, $y_1$). (**b**) The first Poincaré cut $C^s_{E_2}$ and $C^u_{E_2}$ on the plane $y_1$=0, $x_1$<0.

Figure 7(a) shows the projection of the stable manifold tube $M_{E_2}^{s}$ and the unstable manifold tube $M_{E_2}^{u}$ onto the position space ($x_1$, $y_1$). Figure 7(b) shows the first Poincaré cut $C_{E_2}^{s}$ and $C_{E_2}^{u}$ on the plane $y_1$=0, $x_1$<0, and it can be seen that there are twelve points of intersection: two points lie on the $x_1$-axis corresponding to symmetrical homoclinic orbits and ten points lie off the axis corresponding to asymmetrical homoclinic orbits. The leftmost point $H_1$ lies on the $x_1$-axis at about $x_1$=-3.02. The homoclinic orbit corresponding to the point $H_1$ is presented in Fig. 8, and this orbit is symmetrical with respect to the $x_1$-axis. The rightmost point $H_2$ is at about $x_1$=-2.78. The asymmetrical homoclinic orbit corresponding to the point $H_2$ is presented in Fig. 9.

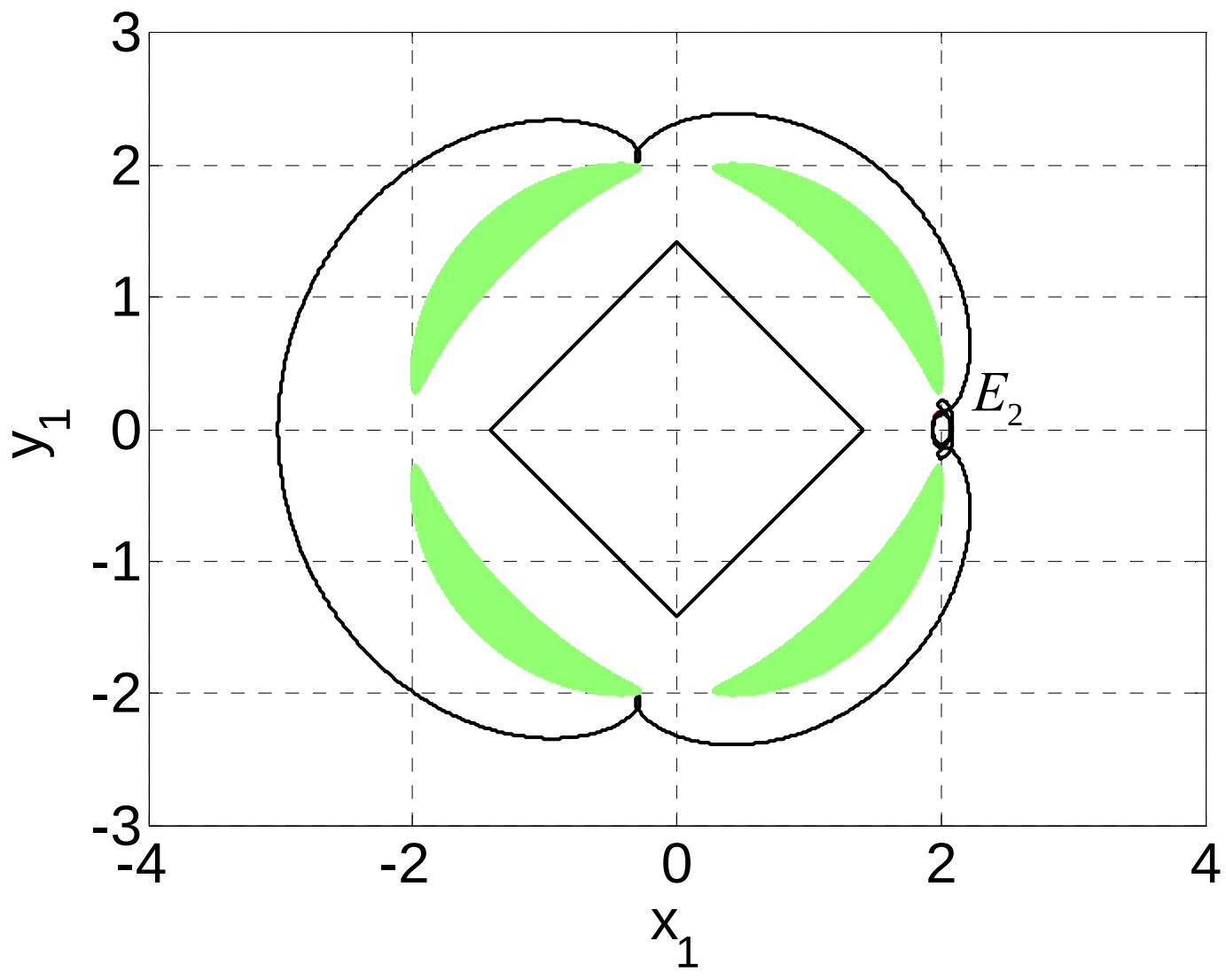

**Fig. 8** Symmetrical homoclinic orbit connecting the $E_2$ periodic orbit in the exterior realm corresponding to the point $H_1$.

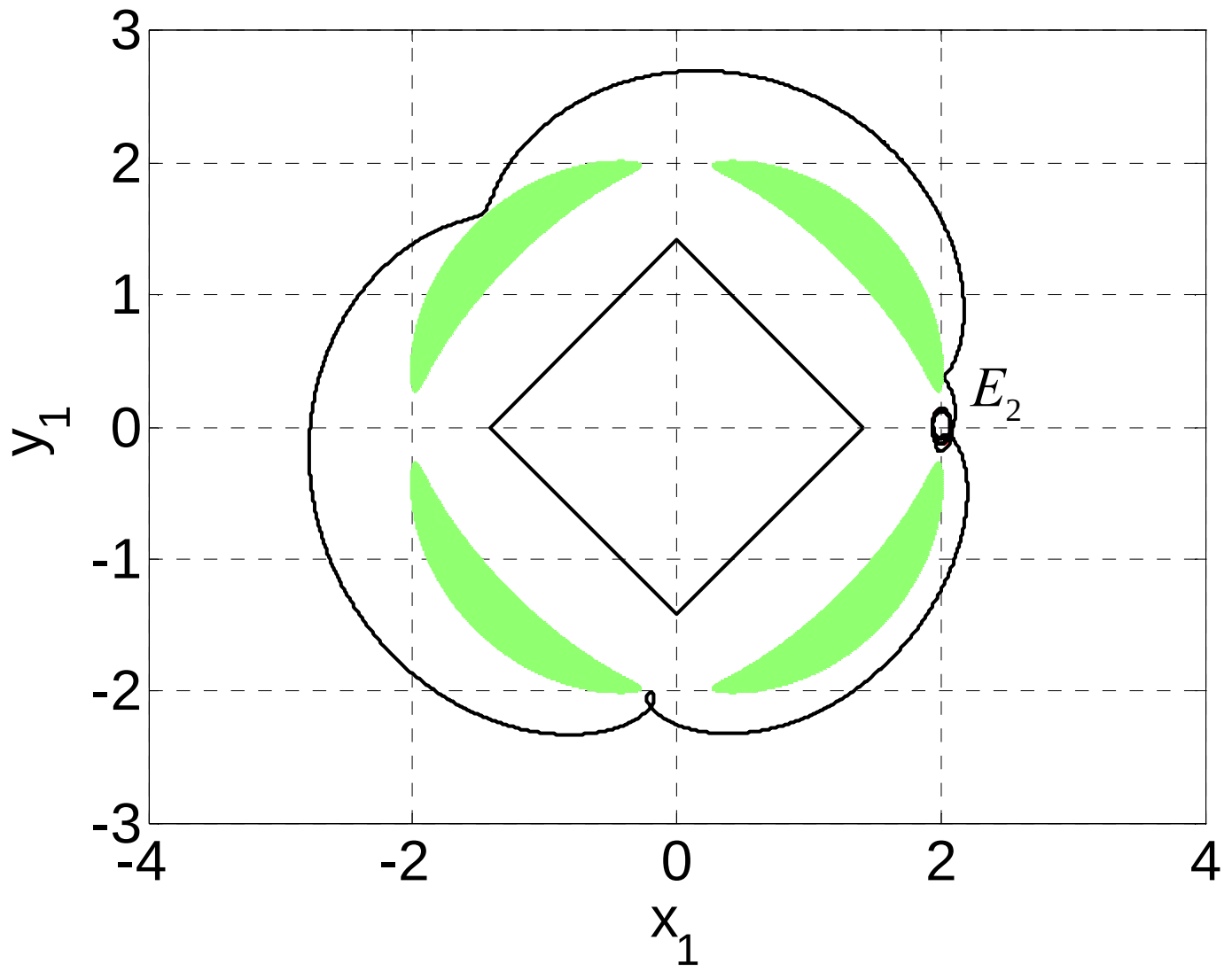

**Fig. 9** Asymmetrical homoclinic orbit connecting the $E_2$ periodic orbit in the exterior realm corresponding to the point $H_2$.

### 6.2 Heteroclinic orbits connecting periodic orbits in the exterior realm ($E_2$—$E_6$)

Heteroclinic orbits connecting periodic orbits can be obtained by finding an intersection of their respective invariant manifolds. In this case, the heteroclinic orbit connecting the $E_2$ and $E_6$ periodic orbits starts from the $E_6$ periodic orbit when $t \to -\infty$, and arrives at the $E_2$ periodic orbit when $t \to +\infty$. Therefore, the stable manifold $M^s_{E_2}$ of the $E_2$ periodic orbit and the unstable manifold $M^u_{E_6}$ of the $E_6$ periodic orbit coincide with each other. The amplitudes of the $E_2$ and $E_6$ periodic orbits are set to be equal, so the energies of the two orbits are also equal. The amplitude $A_m$=0.14 is selected for the sake of clarity.

Because the symmetrical characteristic applies here, System (5) is also invariant to the transformations

$$\left(x_1, y_1, \dot{x}_1, \dot{y}_1, t\right) \rightarrow \left(-x_1, y_1, \dot{x}_1, -\dot{y}_1, -t\right) \tag{20}$$

Therefore, only the stable manifold $M_{E_2}^s$ and the stable cut $C_{E_2}^s$ are plotted here because the unstable manifold $M_{E_6}^u$ and the unstable cut $C_{E_6}^u$ are just mirror images of $M_{E_2}^s$ and $C_{E_2}^s$, respectively.

Figure 10(a) shows the projection of the stable manifold tube $M_{E_2}^s$ onto the position space ($x_1$, $y_1$). Figure 10(b) shows the first Poincaré cut $C_{E_2}^s$ of the manifold $M_{E_2}^s$ on the plane $x_1$=0, $y_1$>0. It can be seen that there are two points of intersection, and both of them lie on the $y_1$-axis. The heteroclinic orbit corresponding to the left point $H_3$ is presented in Fig. 11.

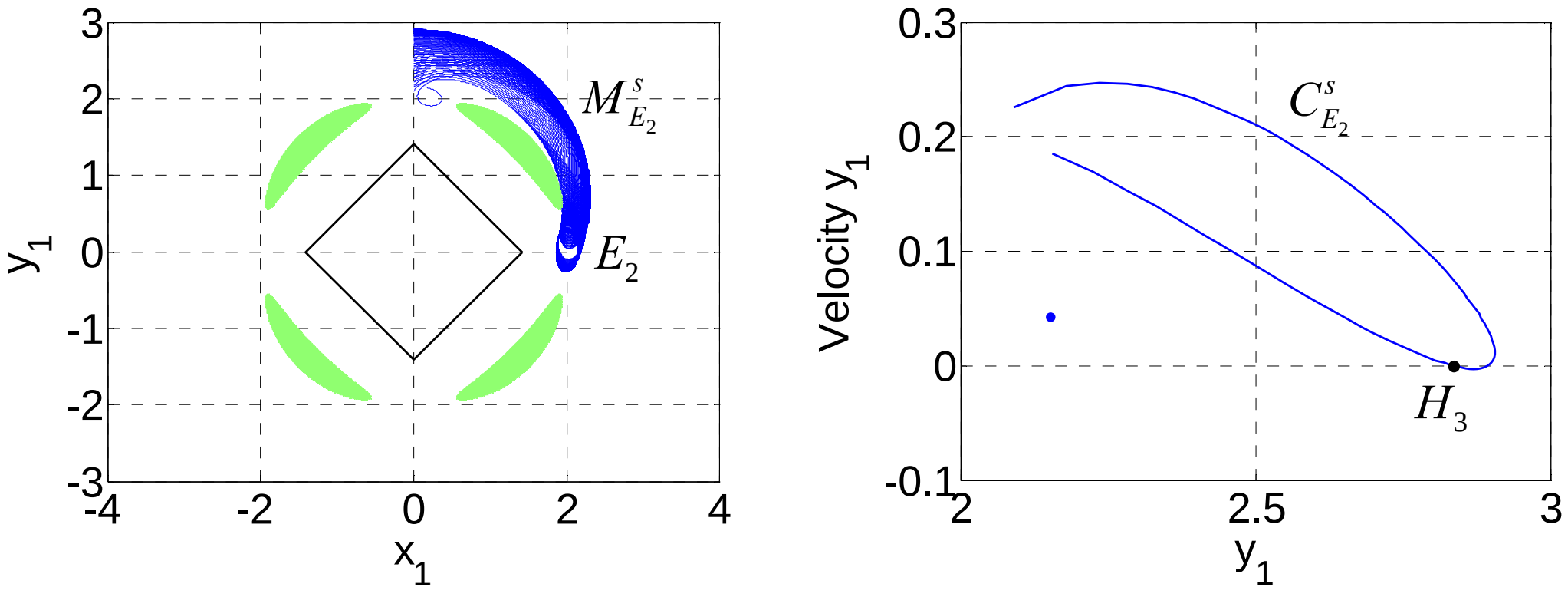

(a) (b)

**Fig. 10** (**a**) Projection of the invariant manifold tube $M_{E_2}^s$ onto the position space ($x_1$, $y_1$). (**b**) The first Poincaré cut $C_{E_2}^s$ of the manifold $M_{E_2}^s$ on the plane $x_1$=0, $y_1$>0.

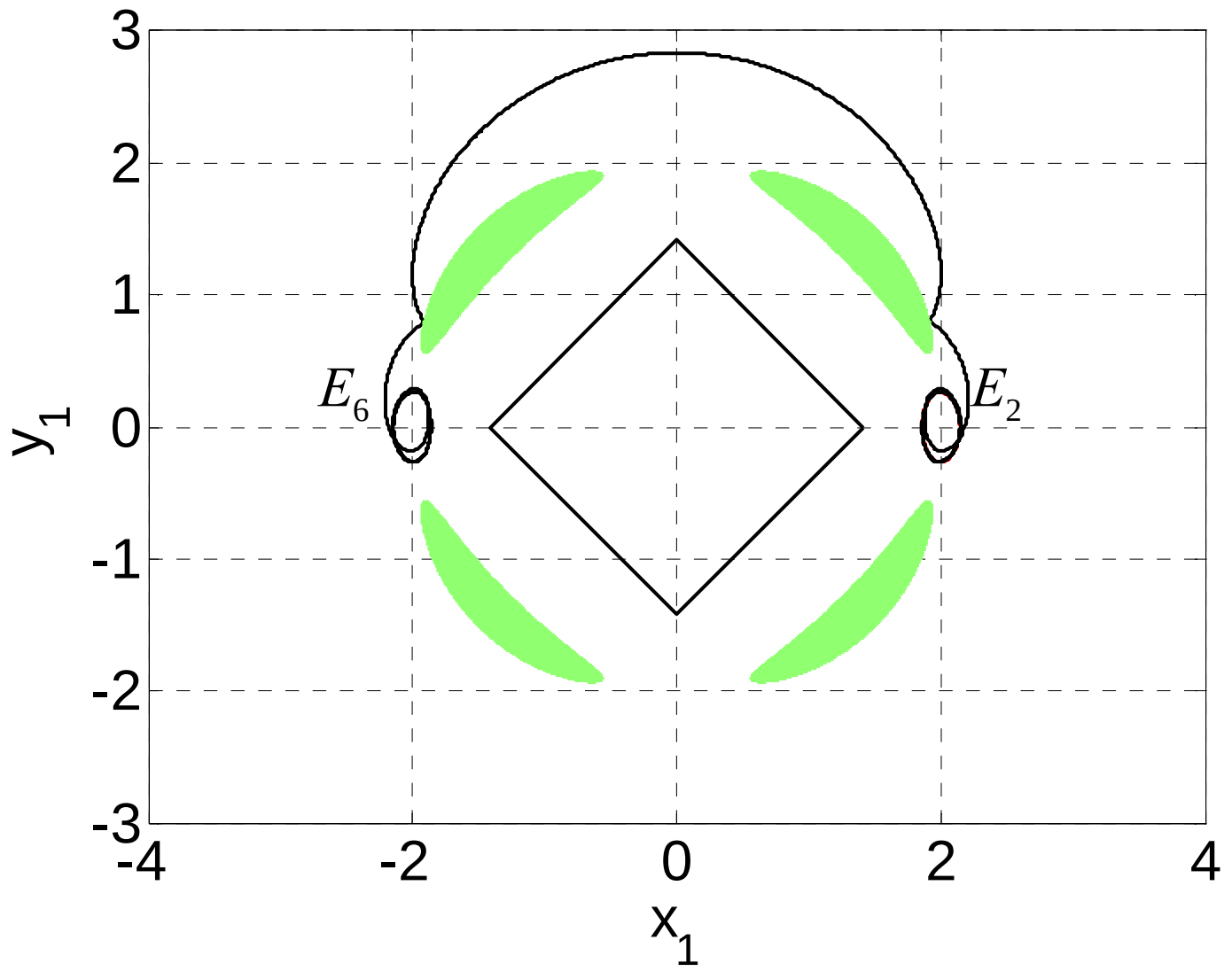

**Fig. 11** Heteroclinic orbit connecting the $E_2$ and $E_6$ periodic orbits in the exterior realm corresponding to the point $H_3$.

**6.3 Heteroclinic orbits connecting periodic orbits in the exterior realm ($E_2$—$E_8$)**

Heteroclinic orbits connecting the $E_2$ and $E_8$ periodic orbits can be obtained by finding the stable manifold $M^s_{E_2}$ of the $E_2$ periodic orbit and the unstable manifold $M^u_{E_8}$ of the $E_8$ periodic orbit. The plane $x_1$ =0, $y_1$>0 is taken as the Poincaré surface of section, so this case is a little different because the symmetrical characteristic does not apply here.

The amplitude $A_m$=0.07 is selected for both the $E_2$ and $E_8$ periodic orbits. Figure 12(a) shows the projection of the stable manifold tube $M^s_{E_2}$ and the unstable manifold tube $M^u_{E_8}$ onto the position space ($x_1$, $y_1$). Figure 12(b) shows the first Poincaré cut $C^s_{E_2}$ and $C^u_{E_8}$ on the plane $x_1$ =0, $y_1$>0. It can be seen that there are four points of intersection, and they all lie off the axis. The heteroclinic orbit corresponding to the point $H_4$ is presented in Fig. 13.

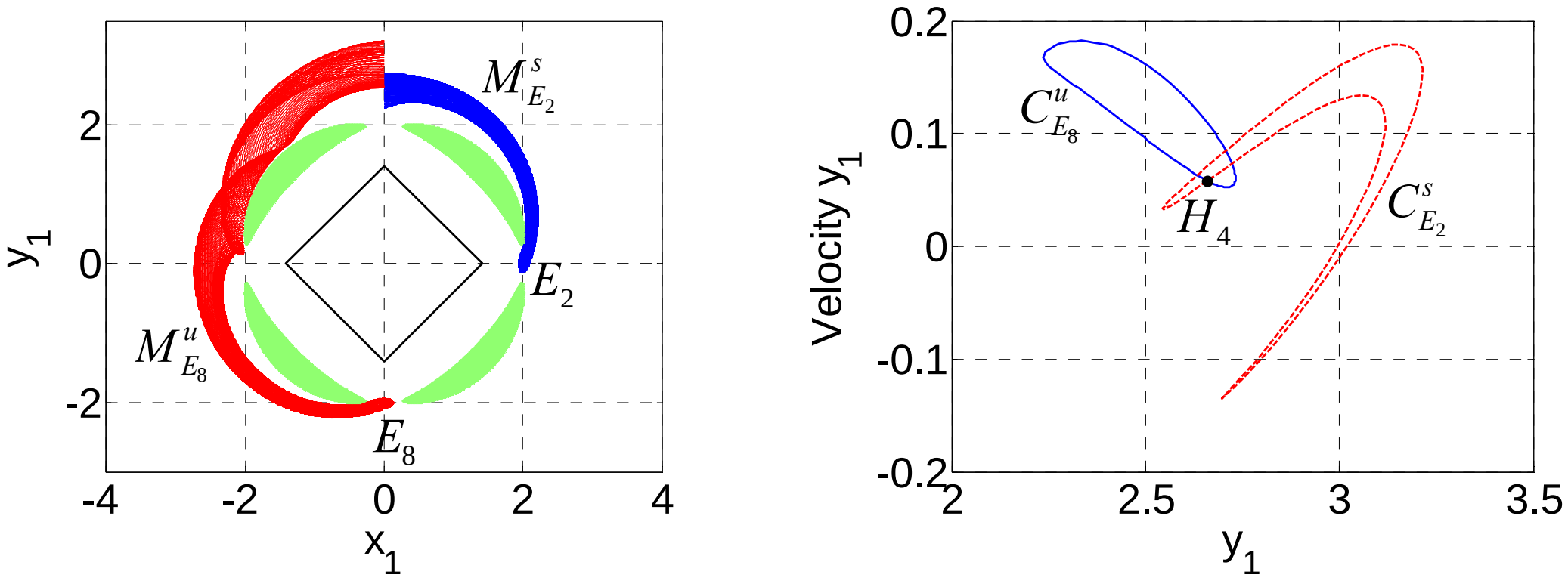

**(a)** **(b)**

**Fig**. **12** (**a**) Projection of the invariant manifold tubes $M^s_{E_2}$ and $M^u_{E_8}$ onto the position space ($x_1$, $y_1$). (**b**) The first Poincaré cut $C^s_{E_2}$ and $C^u_{E_8}$ on the plane $x_1$ =0, $y_1$>0.

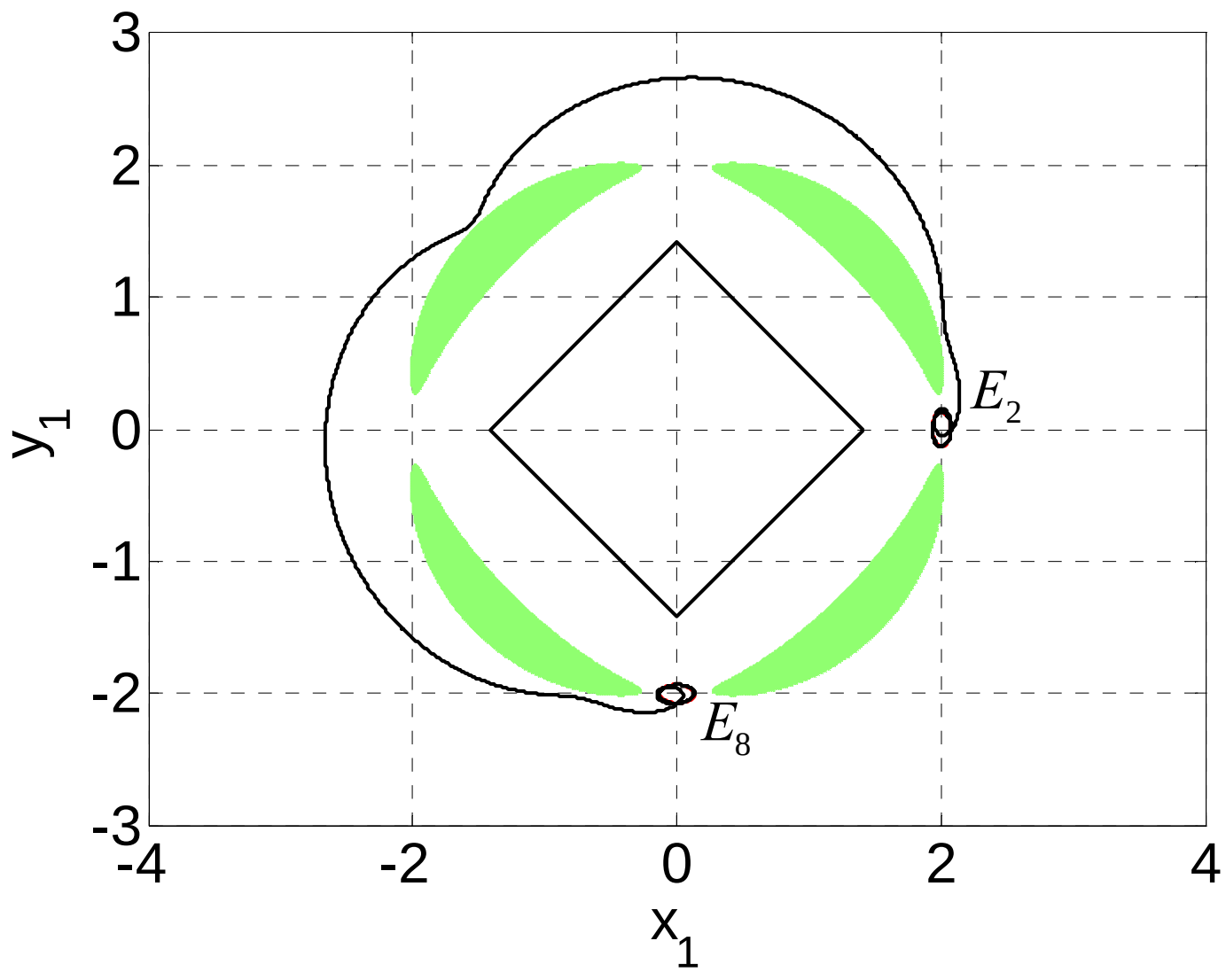

**Fig**. **13** Heteroclinic orbit connecting the $E_2$ and $E_8$ periodic orbits in the exterior realm corresponding to the point $H_4$.

### 6.4 Heteroclinic orbits connecting periodic orbits in the exterior realm ($E_2$—$E_4$)

Heteroclinic orbits connecting the $E_2$ and $E_4$ periodic orbits can also be obtained by finding an intersection of their respective invariant manifolds.

The plane $x$=0, $y$>0 is taken as the Poincaré surface of section, and the amplitude $A_m$=0.1 is selected for both the $E_2$ and $E_4$ periodic orbits.

Figure 14(a) shows the projection of the stable manifold tube $M^s_{E_2}$ and the unstable manifold tube $M^u_{E_4}$ onto the position space ($x_1$, $y_1$). Figure 14(b) shows the first Poincaré cut $C^s_{E_2}$ and $C^u_{E_4}$ on the plane $x$=0, $y$>0. It can be seen that there are two points of intersection, and both lie on the $x$-axis. The heteroclinic orbit corresponding to the point $H_5$ is presented in Fig. 15.

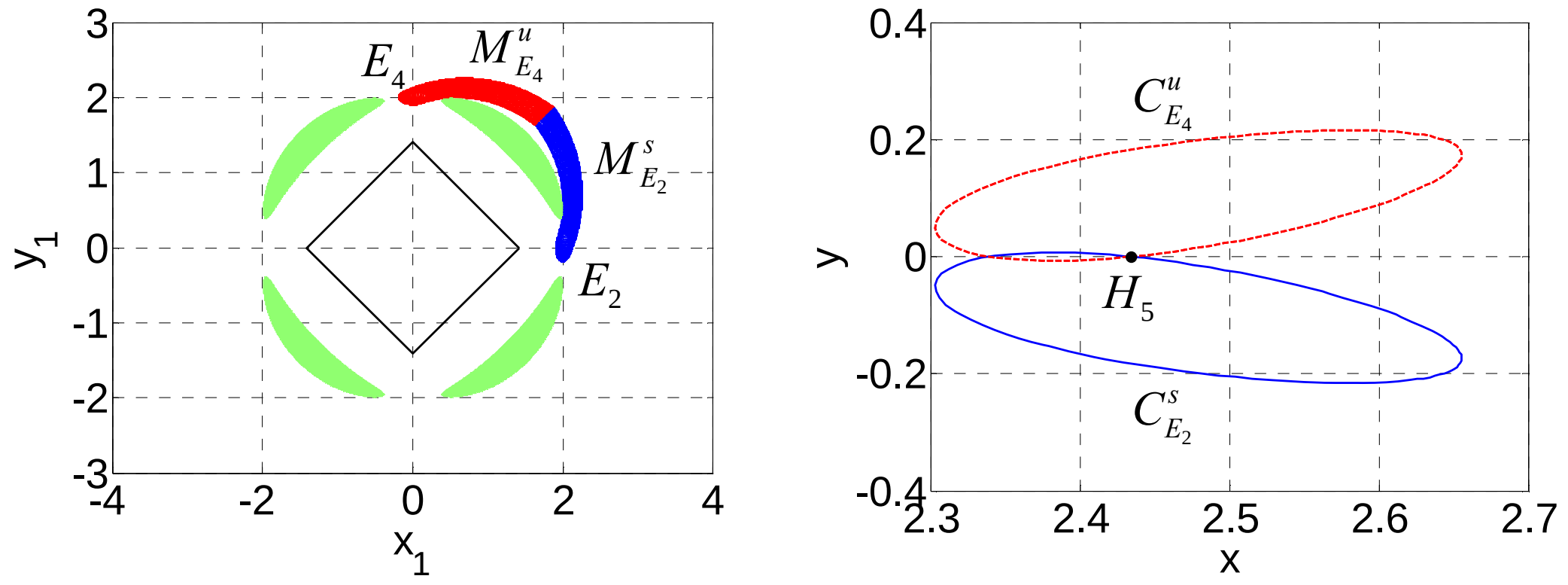

**Fig**. 14 (**a**) Projection of the invariant manifold tubes $M^s_{E_2}$ and $M^u_{E_4}$ onto the position space ($x_1$, $y_1$). (**b**) The first Poincaré cut $C^s_{E_2}$ and $C^u_{E_4}$ on the plane $x$=0, $y$>0.

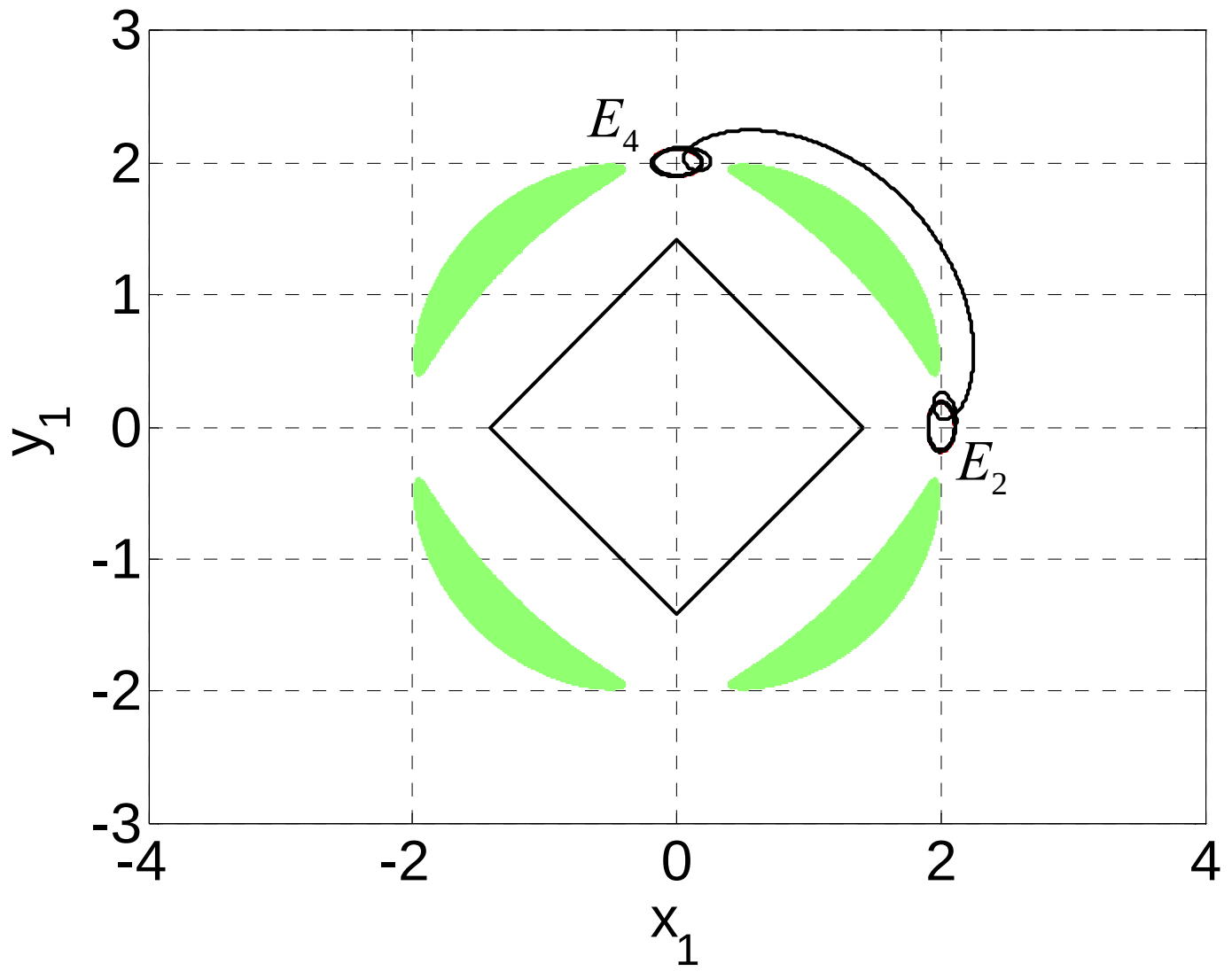

**Fig**. 15 Heteroclinic orbit connecting the $E_2$ and $E_4$ periodic orbits in the exterior realm corresponding to the point $H_5$.

### 6.5 Homoclinic and heteroclinic orbits connecting periodic orbits in the interior realm

Following a similar process, the homoclinic and heteroclinic orbits connecting periodic orbits in the interior realm can be easily obtained, which are shown in Fig. 16.

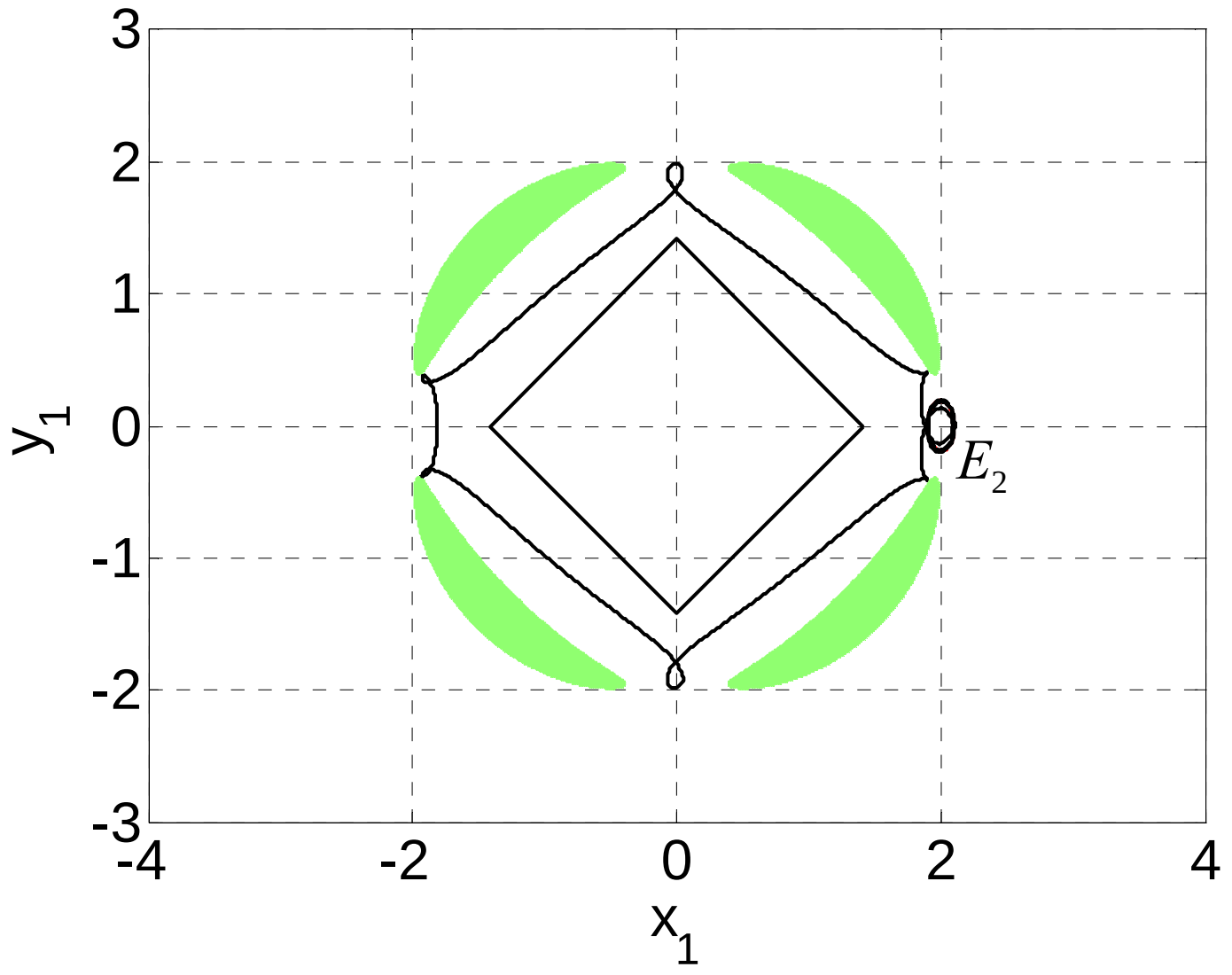

(**a**)

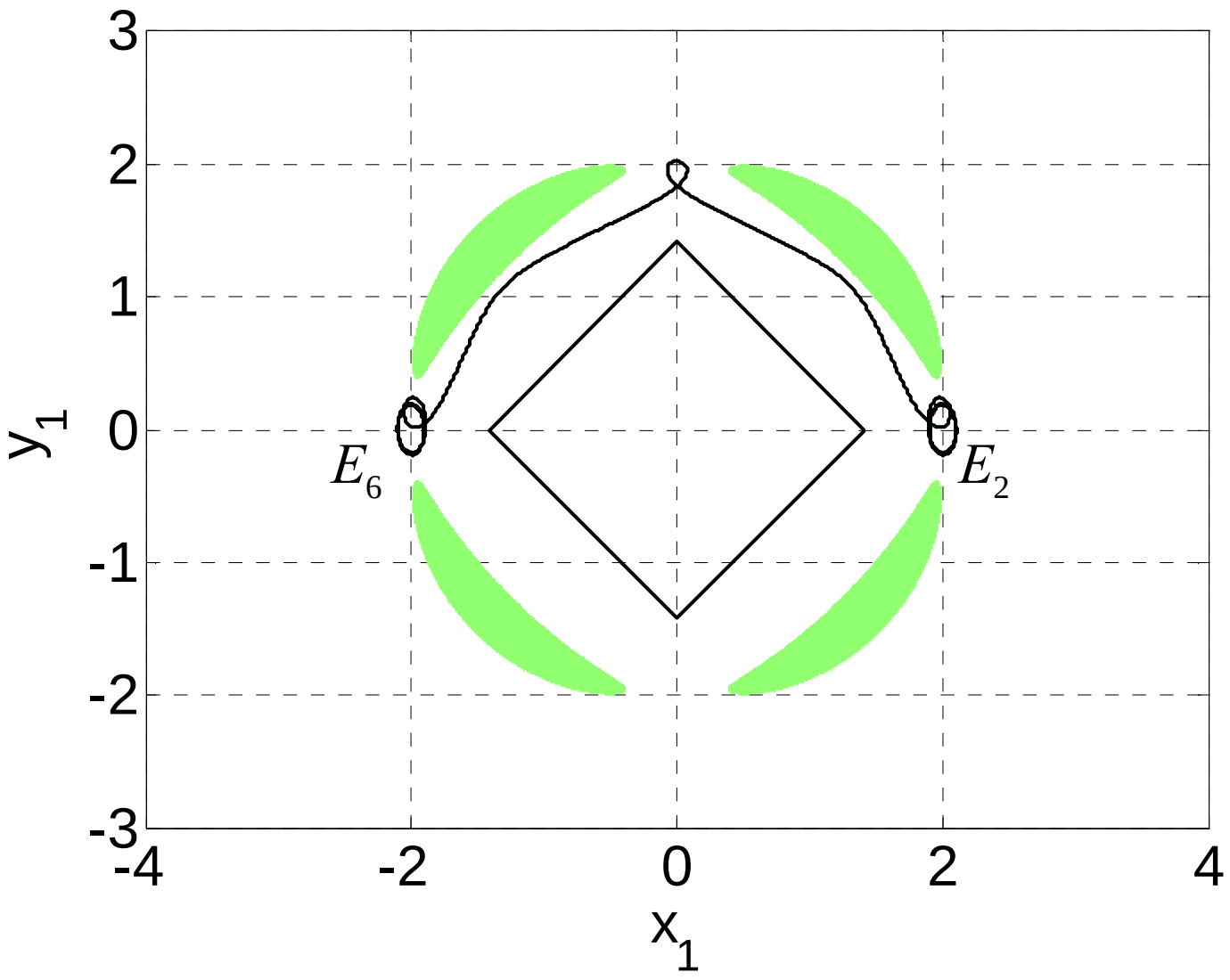

(**b**)

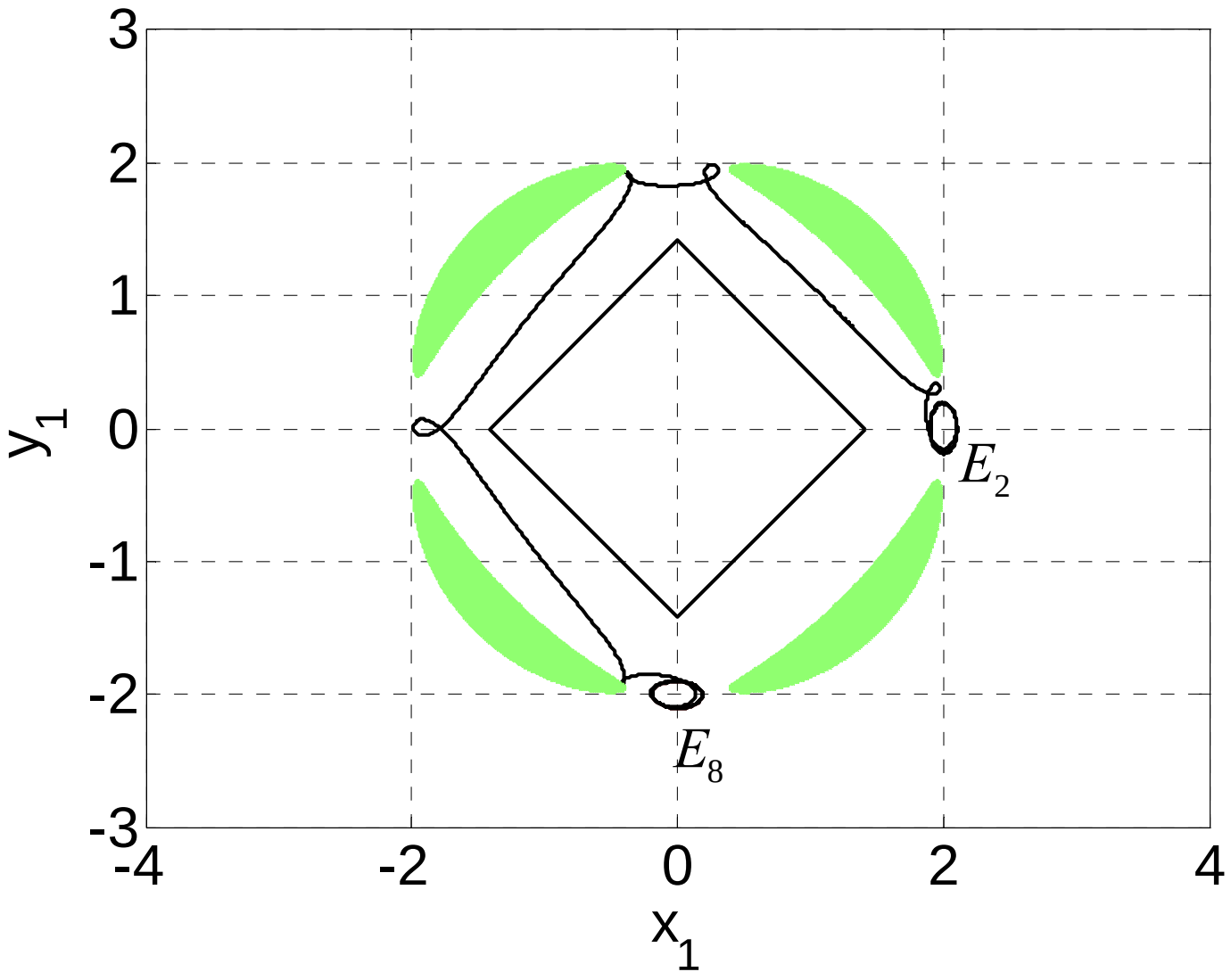

(**c**)

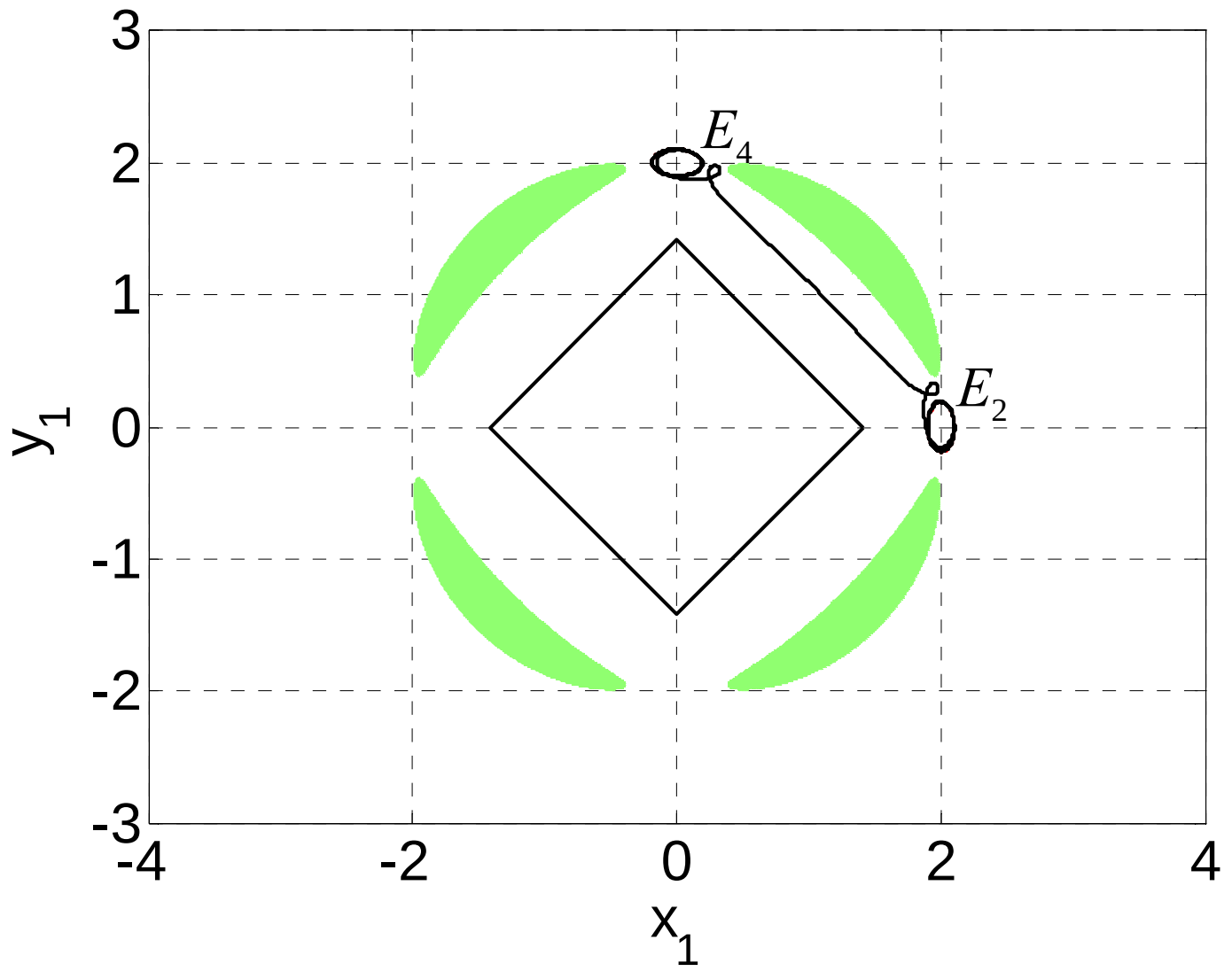

(**d**)

**Fig. 16** Homoclinic and heteroclinic orbits in the interior realm for $A_m$=0.1. (**a**) Homoclinic orbit connecting the $E_2$ periodic orbit. (**b**) Heteroclinic orbit connecting the $E_2$ and $E_6$ periodic orbits. (**c**) Heteroclinic orbit connecting the $E_2$ and $E_8$ periodic orbits. (**d**) Heteroclinic orbit connecting the $E_2$ and $E_4$ periodic orbits.

## 7. Conclusions

This study investigates the dynamics of a particle orbiting around a rotating homogeneous cube, and the results are significant. In the rotating frame of reference, eight equilibria are found, four of which are linearly stable and the other four unstable. Periodic orbits around the equilibria are obtained using the traditional differential correction method. The stabilities of these periodic orbits are determined by the eigenvalues of the monodromy matrix. It is concluded that periodic orbits around the unstable equilibria are also unstable, while periodic orbits around the linearly stable equilibria are stable with small amplitudes. Finally, the existence of the homoclinic and heteroclinic orbits connecting periodic orbits around the equilibria is proved numerically, and some examples are presented.

## Appendix: The potential of the cube

Based on the polyhedral method (Werner and Scheeres 1997), the potential of the fixed cube at a certain point $P$ ($x$, $y$, $z$) in space is calculated as

$$
\begin{aligned}
U = G\sigma\Bigg\{ & -2(x+a)(z-a)\ln\frac{r_1+r_4+2a}{r_1+r_4-2a} + 2(x+a)(y+a)\ln\frac{r_1+r_5+2a}{r_1+r_5-2a} \\
& -2(x-a)(y+a)\ln\frac{r_2+r_6+2a}{r_2+r_6-2a} + 2(x-a)(y-a)\ln\frac{r_3+r_7+2a}{r_3+r_7-2a} \\
& -2(y+a)(z-a)\ln\frac{r_1+r_2+2a}{r_1+r_2-2a} + 2(x-a)(z-a)\ln\frac{r_2+r_3+2a}{r_2+r_3-2a} \\
& +2(y-a)(z-a)\ln\frac{r_3+r_4+2a}{r_3+r_4-2a} - 2(x+a)(y-a)\ln\frac{r_4+r_8+2a}{r_4+r_8-2a} \\
& +2(y+a)(z+a)\ln\frac{r_5+r_6+2a}{r_5+r_6-2a} - 2(x-a)(z+a)\ln\frac{r_6+r_7+2a}{r_6+r_7-2a} \\
& -2(y-a)(z+a)\ln\frac{r_7+r_8+2a}{r_7+r_8-2a} + 2(x+a)(z+a)\ln\frac{r_5+r_8+2a}{r_5+r_8-2a} \\
& +2(a-z)^2\left[\arctan\frac{\mathbf{r}_1\bullet\mathbf{r}_2\times\mathbf{r}_3}{r_1r_2r_3+r_1(\mathbf{r}_2\bullet\mathbf{r}_3)+r_2(\mathbf{r}_1\bullet\mathbf{r}_3)+r_3(\mathbf{r}_1\bullet\mathbf{r}_2)} + \arctan\frac{\mathbf{r}_1\bullet\mathbf{r}_3\times\mathbf{r}_4}{r_1r_3r_4+r_1(\mathbf{r}_3\bullet\mathbf{r}_4)+r_3(\mathbf{r}_1\bullet\mathbf{r}_4)+r_4(\mathbf{r}_1\bullet\mathbf{r}_3)}\right] \\
& +2(a+z)^2\left[\arctan\frac{\mathbf{r}_8\bullet\mathbf{r}_7\times\mathbf{r}_6}{r_6r_7r_8+r_6(\mathbf{r}_7\bullet\mathbf{r}_8)+r_7(\mathbf{r}_6\bullet\mathbf{r}_8)+r_8(\mathbf{r}_6\bullet\mathbf{r}_7)} + \arctan\frac{\mathbf{r}_6\bullet\mathbf{r}_5\times\mathbf{r}_8}{r_5r_6r_8+r_5(\mathbf{r}_6\bullet\mathbf{r}_8)+r_6(\mathbf{r}_5\bullet\mathbf{r}_8)+r_8(\mathbf{r}_6\bullet\mathbf{r}_5)}\right] \\
& +2(a-x)^2\left[\arctan\frac{\mathbf{r}_2\bullet\mathbf{r}_6\times\mathbf{r}_3}{r_2r_3r_6+r_2(\mathbf{r}_3\bullet\mathbf{r}_6)+r_6(\mathbf{r}_2\bullet\mathbf{r}_3)+r_3(\mathbf{r}_2\bullet\mathbf{r}_6)} + \arctan\frac{\mathbf{r}_6\bullet\mathbf{r}_7\times\mathbf{r}_3}{r_3r_6r_7+r_6(\mathbf{r}_7\bullet\mathbf{r}_3)+r_7(\mathbf{r}_6\bullet\mathbf{r}_3)+r_3(\mathbf{r}_6\bullet\mathbf{r}_7)}\right] \\
& +2(a+x)^2\left[\arctan\frac{\mathbf{r}_1\bullet\mathbf{r}_4\times\mathbf{r}_5}{r_1r_4r_5+r_1(\mathbf{r}_4\bullet\mathbf{r}_5)+r_4(\mathbf{r}_1\bullet\mathbf{r}_5)+r_5(\mathbf{r}_1\bullet\mathbf{r}_4)} + \arctan\frac{\mathbf{r}_4\bullet\mathbf{r}_8\times\mathbf{r}_5}{r_4r_5r_8+r_4(\mathbf{r}_5\bullet\mathbf{r}_8)+r_8(\mathbf{r}_4\bullet\mathbf{r}_5)+r_5(\mathbf{r}_4\bullet\mathbf{r}_8)}\right] \\
& +2(a-y)^2\left[\arctan\frac{\mathbf{r}_4\bullet\mathbf{r}_3\times\mathbf{r}_7}{r_3r_4r_7+r_3(\mathbf{r}_4\bullet\mathbf{r}_7)+r_4(\mathbf{r}_3\bullet\mathbf{r}_7)+r_7(\mathbf{r}_3\bullet\mathbf{r}_4)} + \arctan\frac{\mathbf{r}_7\bullet\mathbf{r}_8\times\mathbf{r}_4}{r_4r_7r_8+r_4(\mathbf{r}_7\bullet\mathbf{r}_8)+r_7(\mathbf{r}_4\bullet\mathbf{r}_8)+r_8(\mathbf{r}_4\bullet\mathbf{r}_7)}\right] \\
& +2(a+y)^2\left[\arctan\frac{\mathbf{r}_1\bullet\mathbf{r}_5\times\mathbf{r}_6}{r_1r_5r_6+r_1(\mathbf{r}_5\bullet\mathbf{r}_6)+r_5(\mathbf{r}_1\bullet\mathbf{r}_6)+r_6(\mathbf{r}_1\bullet\mathbf{r}_5)} + \arctan\frac{\mathbf{r}_2\bullet\mathbf{r}_1\times\mathbf{r}_6}{r_1r_2r_6+r_1(\mathbf{r}_2\bullet\mathbf{r}_6)+r_2(\mathbf{r}_1\bullet\mathbf{r}_6)+r_6(\mathbf{r}_1\bullet\mathbf{r}_2)}\right]\Bigg\},
\end{aligned}
$$

where the Cartesian coordinate system *Oxyz* is established with the origin *O* located at the center of the fixed cube and the three coordinate axes coinciding with the symmetrical axes of the cube; $G$ is the gravitational constant; $\sigma$ is the constant mass density of the cube; $a$ is the half-length of the cubic edge; $\mathbf{r}_i$ ($i$=1,2,…,8) is the vector from the origin *O* to one of the eight vertices of the cube, and $r_i$ is the norm of $\mathbf{r}_i$.

## Acknowledgments

This work was supported by the National Natural Science Foundation of China (No. 10832004 and No. 11072122).